\documentclass[12pt]{iopart}
\usepackage{iopams}
\usepackage{graphicx}
\begin{document}

\title[Evolution from localized to intermediate valence regime in Ce$_2$Cu$_{2-x}$Ni$_x$In]{Evolution from localized to intermediate valence regime in Ce$_2$Cu$_{2-x}$Ni$_x$In}

\author{A~P~Pikul and D Kaczorowski}

\address{Institute of Low Temperature and Structure Research, Polish Academy of Sciences, P Nr 1410, 50--590~Wroc{\l}aw 2, Poland}
\ead{A.Pikul@int.pan.wroc.pl}

\begin{abstract}
Polycrystalline samples of the solid solution Ce$_2$Cu$_{2-x}$Ni$_x$In were studied by means of
x-ray powder diffraction, magnetic susceptibility and electrical resistivity measurements performed
in a wide temperature range. Partial substitution of copper atoms by nickel atoms results in
quasi-linear decrease of the lattice parameters and the unit cell volume of the system. The lattice
compression leads to an increase in the exchange integral and yields a reversal in the order of the
magnetic $4f^1$ and nonmagnetic $4f^0$ states, being in line with the Doniach phase diagram. In the
localized regime, where an interplay of the Kondo scattering and the crystalline electric field
effect takes place, the rise in the hybridization strength is accompanied with relative reduction
in the scattering conduction electrons on excited crystal field levels.
\end{abstract}


\submitto{\JPCM}

%\maketitle

\section{Introduction}
Ternary intermetallics Ce$_2$Cu$_2$In and Ce$_2$Ni$_2$In crystallize in a~primitive tetragonal
structure of the Mo$_2$FeB$_2$ type (space group $P4/mbm$) with the lattice parameters
$a$~=~7.7368(6)~\AA, $c$~=~3.9240(3)~\AA~ for the former phase, and $a$~=~7.5305(3)~\AA,
$c$~=~3.7223(2)~\AA~ for the latter one~\cite{kaczorowski}. The physical properties of
Ce$_2$Cu$_2$In indicate well localized magnetism due to the presence of stable Ce$^{3+}$ ions. The
compound orders antiferromagnetically at the N\'{e}el temperature $T_{\rm N}$~=~5.5~K, and its
electrical resistivity is dominated by an interplay of strong Kondo and crystalline electric field
interactions. In contrast, the Ce$_2$Ni$_2$In compound exhibits features characteristic of
intermediate valence systems, with partly delocalized $4f$ electrons of
cerium~\cite{kaczorowski,hauser}.

Since both compounds are isostructural, the qualitative difference in their ground states results
exclusively from swapping copper atoms for nickel atoms. This fact motivated us to perform alloying
studies of Ce$_2$Cu$_{2-x}$Ni$_x$In as a unique opportunity to investigate an evolution of a $4f$
system from fully localized regime to well defined intermediate valence state, that is predicted by
the Doniach diagram \cite{doniach}. It is worth noting that systematic studies of such evolution
are often difficult (or even impossible) due to either relatively small difference between the
ground states of two isostructural parent compounds or different crystal structures of the two
terminal phases. In other words, such experimental studies usually cover only part of the Doniach
diagram, e.g. quantum critical region.

In this paper we present the results of magnetic and electrical transport measurements of the solid
solution Ce$_2$Cu$_{2-x}$Ni$_x$In, performed in a wide temperature range, using polycrystalline
samples. We demonstrate a prominent evolution of the main physical characteristics of the system
upon increasing nickel content, and interpret the experimental data in terms of some theoretical
models developed for localized and intermediate valence states. We argue, that the Cu/Ni
substitution leads to an increase in the exchange integral, which yields a reversal in the order of
the magnetic $4f1$ and nonmagnetic $4f0$ states. Moreover, the rise in the exchange integral brings
about a diminishing influence of excited crystal field levels on the electrical transport in the
system studied.

\section{Experimental details}
Polycrystalline samples of Ce$_2$Cu$_{2-x}$Ni$_x$In were prepared by arc melting stoichiometric
amounts of the elemental components in titanium gettered argon atmosphere. The pellets were turned
over and remelted several times to ensure good homogeneity. No further heat treatment was given to
the as cast ingots.

Quality of the products was verified by X-ray powder diffraction (Stoe diffractometer with Cu
$K{\alpha}_1$ radiation) and energy dispersive X-ray spectroscopy (Phillips 515 scanning electron
microscope equipped with EDAX PV 9800 spectrometer). All the diffractograms were easily indexed
within the expected primitive tetragonal structure. The microprobe analysis showed that the samples
are nearly single phase (with minor traces of unidentified impurities) and their compositions are
close to the nominal ones.

The magnetic properties were studied at temperatures 1.7~--~400~K and in magnetic fields up to 5~T
using a commercial Quantum Design SQUID magnetometer. Electrical resistivity was measured on
bar-shaped samples in zero field from 4.2~K up to room temperature, using a home-made setup.

\section{Results and discussion}

\subsection{Crystal structure}

\begin{figure}
\centering %%
\includegraphics[width=7cm]{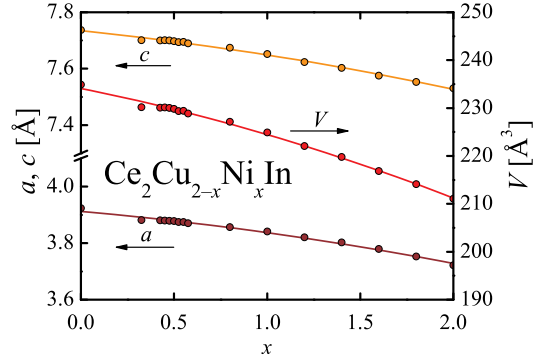}%
\caption{\label{fig_lattice_param}  Lattice parameters $a$ and $c$ and unit cell volume $V$ of selected alloys Ce$_2$Cu$_{2-x}$Ni$_x$In as a function of the nominal nickel content $x$.}
\end{figure}

Analysis of the X-ray diffraction patterns (not given here) confirmed that substitution of copper
atoms by nickel atoms does not change the crystal structure of the system. It revealed also that
the lattice parameters and the unit cell volume of the Ce$_2$Cu$_{2-x}$Ni$_x$In alloys
systematically decrease (in a quasi-linear manner) with increasing nickel content from
$a$~=~7.7368(6)~\AA, $c$~=~3.9240(3)~\AA~ and $V$~=~234.88~\AA$^3$ for Ce$_2$Cu$_2$In to
$a$~=~7.5305(3)~\AA, $c$~=~3.7223(2)~\AA~ and $V$~=~211.09~\AA$^3$ for
Ce$_2$Cu$_2$In~\cite{kaczorowski}  (see \fref{fig_lattice_param}).

\subsection{Magnetic properties}

\begin{figure*}
\centering %%
\includegraphics[width=14cm]{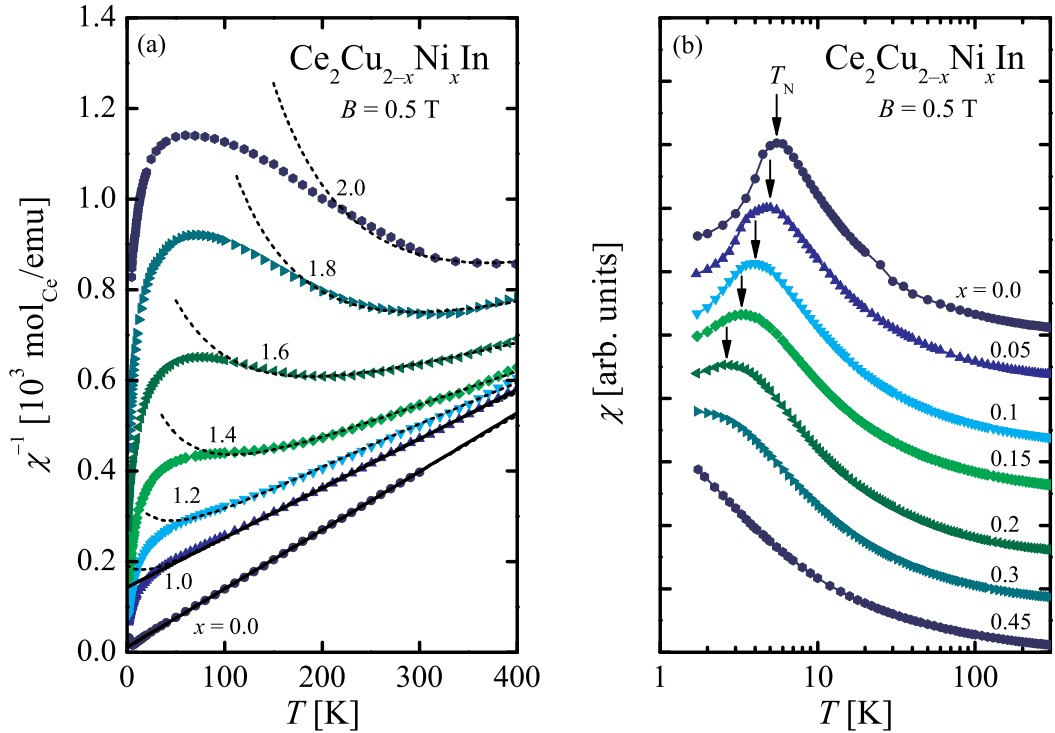}%
\caption{\label{fig_suscept}  (a) Inverse magnetic susceptibility $\chi^{-1}$ of selected alloys
Ce$_2$Cu$_{2-x}$Ni$_x$In as a function of temperature $T$. Straight solid lines represent
Curie-Weiss fittings; results for $0.00 < x < 1.00$ are omitted for the sake of clarity. (b)
Magnetic susceptibility $\chi$ of Ce$_2$Cu$_{2-x}$Ni$_x$In with $x \leqslant 0.45$ vs. $T$. Solid
curves serve as guides for the eye and the arrows mark the N{\'{e}}el temperature $T_{\rm N}$. $x$
is the nominal composition.}
\end{figure*}

\begin{figure}
\centering %%
\includegraphics[width=7cm]{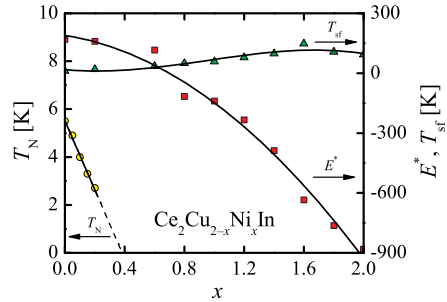}%
\caption{\label{fig_diagram}  N{\'e}el temperature $T_{\rm N}$, excitation energy $E^{\ast}$ and spin fluctuation temperature $T_{\rm sf}$ of Ce$_2$Cu$_{2-x}$Ni$_x$In as a function of the nominal nickel content $x$. Solid and dashed lines serve as guides for the eye.}
\end{figure}

\Fref{fig_suscept}(a) presents the temperature dependencies of the inverse magnetic susceptibility
of the alloys studied. As seen, $\chi^{-1}(T)$ measured for the Cu rich alloys (i.e. for $x
\leqslant 1.00$) exhibit above about 100~K a linear behaviour given by the conventional
Curie--Weiss law:
\begin{equation}
\label{cw}
\chi (T) = \frac{1}{8} \frac{\mu_{\rm eff}^2}{T-\theta_{\rm p}},
\end{equation}
where $\mu_{\rm eff}$ is the effective magnetic moment and $\theta_{\rm p}$ stands for the Weiss
temperature. Least squares fittings of \eref{cw} to the experimental data collected for the alloys
with $x \leqslant 1.00$ [see the solid lines in \fref{fig_suscept}(a)] yielded nearly constant
$\mu_{\rm eff}$ of about 2.5--2.6~$\mu_{\rm B}$ and $\theta_{\rm p}$ decreasing monotonically from
about $-16$~K for Ce$_2$Cu$_2$In (in agreement with \cite{kaczorowski}) to $-51$~K for
Ce$_2$Cu$_{1.0}$Ni$_{1.0}$In. The obtained values of the effective magnetic moments are close to
the theoretical one calculated for free Ce$^{3+}$ ions (i.e. 2.54~$\mu_{\rm B}$) and point at the
presence of well localized $4f$ electrons of cerium. The increase in the absolute value of the
Weiss temperature can be ascribed to the enhancement of the hybridization strength between
conduction electrons and $4f$ shells, expected for rising the Ni content.

The latter hypothesis is in line with further evolution of $\chi^{-1}(T)$, that is observed for the
alloys with $x > 1$. Along with increasing the nickel content, the inverse magnetic susceptibility
deviates more and more from the Curie--Weiss behaviour, suggesting progressive delocalization of
the $4f$ electrons (i.e. evolution from the state $4f^1$ to the configuration $4f^0$).

According to the interconfiguration fluctuation (ICF) model developed by Sales and
Wohlleben~\cite{sales}, the magnetic susceptibility of the $4f$ electron compounds can be described
as a sum of susceptibilities of two states: $4f^n$ and $4f^{n-1}$. Respectively, the states are
characterized by well defined angular momenta $J_n$ and $J_{n-1}$, the degeneracies $2J_n +1$ and
$2J_{n-1} +1$, the Hund's-rule effective magnetic moments $\mu_n$ and $\mu_{n-1}$, and the energies
$E_n$ and $E_{n-1}$. The total magnetic susceptibility $\chi(T)$ of such a system can be expressed
as:
\begin{equation}
\label{chi_sw}
\chi(T) = \frac{1}{8}\frac{\nu(T) \mu_n^2}{T  + T_{\rm sf}}
+ \frac{1}{8}\frac{\left[ 1- \nu(T) \right]\mu_{n-1}^2}{T  + T_{\rm sf}},
\end{equation}
with the fractional occupation $\nu(T)$ of the $4f^n$ state:
\begin{equation}
\label{nu_sw}
\nu (T) = \frac{(2J_n +1)}{(2J_n +1) + (2J_{n-1} +1)\exp{\left[ -E^{\ast}/(T + T_{\rm sf})
\right]}},
\end{equation}
where $T_{\rm sf}$ is the spin-fluctuation temperature related to the transition rate $\omega_f$
between the two electron configurations involved ($k_{\rm B} T_{\rm sf} = \hbar \omega_f$), and
$E^{\ast}$ is the energy difference between the two $4f$ states ($E^{\ast} = E_{n-1} - E_n$),
expressed in kelvins.

In cerium compounds, there are two possible configurations of the $4f$ shell: $4f^1$ (with $J_1 =
\frac{5}{2}$ and $\mu_1$~=~2.54~$\mu_{\rm B}$) and $4f^0$ (with $J_0$~=~0 and $\mu_0$~=~0). Hence,
\eref{chi_sw} and \eref{nu_sw} take the form, respectively:
\begin{equation}
\label{chi_sw_ce}
\chi(T) = \frac{1}{8}\frac{\nu(T) (2.54 \mu_{\rm B})^2}{T  + T_{\rm sf}},
\end{equation}
and:
\begin{equation}
\label{nu_sw_ce}
\nu (T) = \frac{6}{6 + \exp{\left[ -E^{\ast}/(T  + T_{\rm sf}) \right]}},
\end{equation}
with $E^{\ast} = E_0 - E_1$. According to this notation, in systems with the magnetic ground state
$4f^1$ (e.g. magnetically ordered compounds, Kondo and heavy fermion systems) $E^{\ast}$ is
positive, while in the intermediate valence systems, in which the $4f^1$ configuration is the
excited state, $E^{\ast}$ is negative. For $E^{\ast}\geqslant 0$, the fractional occupation $\nu$
of the ground $4f^1$ state is weakly dependent on temperature and decreases from 1 (for $E^{\ast}
\rightarrow +\infty$) to 6/7 (for $E^{\ast}=0$), and the associated magnetic moment changes from
2.54~$\mu_{\rm B}$ to 2.35~$\mu_{\rm B}$, respectively. Thus, the system exhibits features
characteristic of localized magnetism, and \eref{chi_sw_ce} is a good approximation of the
Curie--Weiss law. In turn, for $E^{\ast} < 0$, a characteristic maximum in $\chi(T)$ develops with
increasing $|E^{\ast}|$, yet the Curie--Weiss behaviour can still be observed at high enough
temperatures ($T >> |E^{\ast}|$).

The magnetic susceptibility of the alloys Ce$_2$Cu$_{2-x}$Ni$_x$In seems to nicely follow the
predictions of the ICF model. The experimental data can be described at elevated temperatures by
\eref{chi_sw_ce} modified by adding a constant term $\chi_0$, which accounts for possible
temperature independent paramagnetic and diamagnetic contributions to the total magnetic
susceptibility (cf. \cite{kaczorowski, pecharsky}), i.e.:
\begin{equation}
\label{chi_sw_ce_final}
\chi(T) = \frac{1}{8}\frac{\nu(T) (2.54 \mu_{\rm B})^2}{T  + T_{\rm sf}} + \chi_0.
\end{equation}
Least squares fittings of \eref{chi_sw_ce_final} to the experimental data [see the dashed lines in
\fref{fig_suscept}(a) and \fref{fig_diagram}] yielded $E^{\ast}$ decreasing monotonically from
$+169(18)$~K for Ce$_2$Cu$_2$In to $-882(9)$~K for Ce$_2$Ni$_2$In (in agreement with the value of
--884~K, reported previously~\cite{kaczorowski}). The parameter $T_{\rm sf}$ was found to increase
from +11(1)~K to 93(4)~K, respectively, in a non-linear manner with a local maximum of 147(5)~K for
Ce$_2$Cu$_{0.4}$Ni$_{1.6}$In. In turn, the temperature independent term was found to be nearly
constant in the whole series of the alloys and equal to $\chi_0 \! \sim \! 10^{-4}$~emu/mol.

As apparent from \fref{fig_suscept}(a), at low temperatures the experimental $\chi^{-1}(T)$
variations measured for strongly intermediate valent alloys significantly deviate from the
predictions of the ICF model (note Curie-like tails $\chi\! \sim \! 1/T$). Such a behaviour is
often observed in the intermediate valence systems and can be ascribed to the presence of
paramagnetic impurities (cf. \cite{kaczorowski}). While in the localized regime the latter
contribution is negligible, in the intermediate valence state (exhibiting relatively weak
magnetism) it becomes significant and thus strongly modifies the expected ICF curvature of
$\chi^{-1}(T)$.

The change of $E^{\ast}$ with increasing the nickel content (\fref{fig_diagram}) is in line with
the observed evolution of the magnetic properties of the system Ce$_2$Cu$_{1-x}$Ni$_{x}$In. One
should note however, that the ICF model does not take into account i.a. the crystal electric field
effect and magnetic correlations, which are also expected to play some role in the alloys studied
(cf. \cite{kaczorowski}). Therefore, the derived absolute values of $E^{\ast}$ can differ from the
actual energy difference between the $4f^1$ and $4f^0$ configurations, especially in the localized
regime.

\Fref{fig_suscept}(b) displays the temperature dependencies of the magnetic susceptibility
$\chi(T)$ for the alloys with $x \leqslant 0.45$, which exhibit well localized magnetic moments.
Clearly, the compound Ce$_2$Cu$_2$In orders antiferromagnetically at the N\'{e}el temperature
$T_{\rm N}$~=~5.5~K, in agreement with the previous finding~\cite{kaczorowski}. With increasing
$x$, $T_{\rm N}$ [defined as a maximum on the $\chi(T)$ curve] rapidly decreases. For $x = 0.45$,
hardly any anomaly is visible in $\chi(T)$, at least in the temperature range studied.
Extrapolation of the dependence $T_{\rm N}(x)$ to the absolute zero temperature (see the dashed
line in \fref{fig_diagram}) yields the critical concentration $x_{\rm c}$ of about 0.4. In parallel
with the suppression of $T_{\rm N}$, one observes gradual broadening of the maximum in $\chi(T)$
that manifests the antiferromagnetic phase transition. This finding suggests that the quantum
critical phase transition smears out and hence no quantum critical state could be reached, as
concluded also e.g. in CePd$_{1-x}$Rh$_x$~\cite{sereni,pikul,westerkamp}.

\subsection{Electric properties}

\begin{figure*}
\includegraphics[width=14cm]{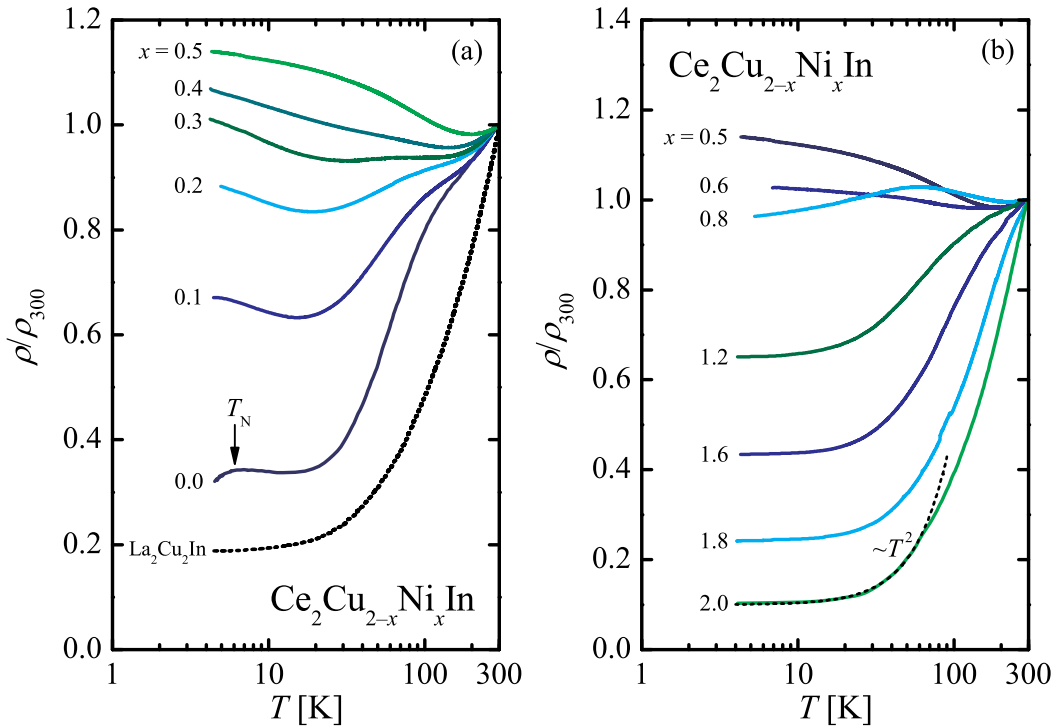}%
\centering %%
\caption{\label{fig_resist}  Electrical resistivity $\rho$ of Ce$_2$Cu$_{2-x}$Ni$_x$In and its
non-magnetic reference La$_2$Cu$_2$In, normalized per room temperature value $\rho_{300}$, as a
function of temperature $T$. $x$ is the nominal composition and the arrow marks the N\'{e}el
temperature $T_{\rm N}$. The data for La$_2$Cu$_2$In are taken from \cite{kaczorowski}.}
\end{figure*}

\begin{figure*}
\includegraphics[width=14cm]{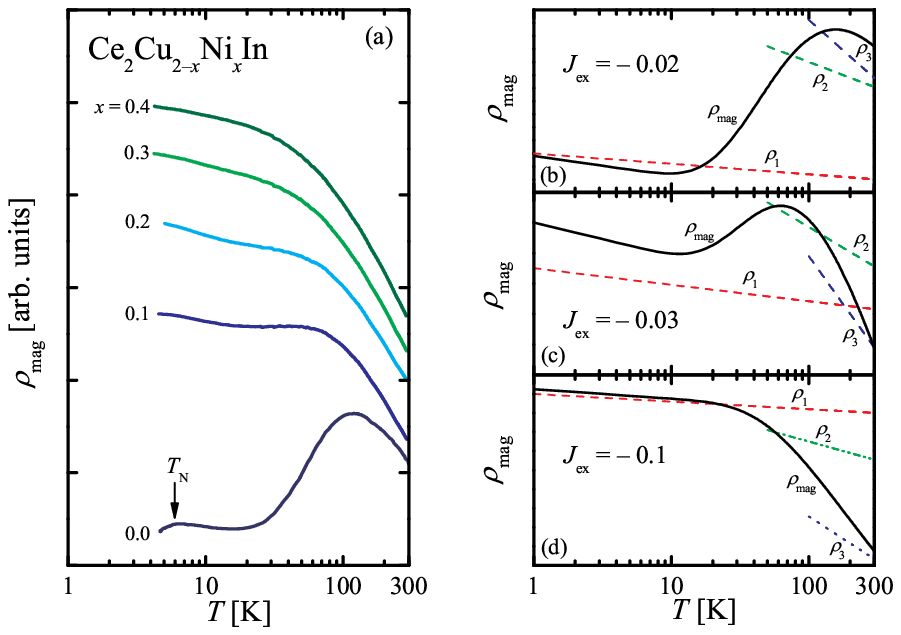}%
\centering %%
\caption{\label{fig_resist_magn}  (a) Temperature dependencies of a Kondo contribution $\rho_{\rm
mag}$ to the electrical resistivity of Ce$_2$Cu$_{2-x}$Ni$_x$In. Panels (b), (c), and (d) display
theoretical $\rho_{\rm mag}(T)$ of a Ce$^{3+}$-based system and tetragonal CEF potential [equation
\eref{cornut_simulation}], calculated for different values of $J_{\rm ex}$. The assumed CEF
splitting scheme: $\Delta_1$~=~0~K, $\Delta_2$~=~50~K and $\Delta_3$~=~100~K. The dashed lines
represent the individual $\rho_i$ contributions [equation \eref{cornut}] for fully populated
crystal field levels.}
\end{figure*}

The temperature variations of the electrical resistivity of Ce$_2$Cu$_{1-x}$Ni$_{x}$In and the
nonmagnetic reference La$_2$Cu$_2$In (normalized to room temperature values) are presented in
\fref{fig_resist}. While the $\rho(T)$ curve obtained for the La sample is characteristic of simple
metals, the resistivity of the parent compound Ce$_2$Cu$_2$In exhibits features of a magnetically
ordered dense Kondo system. In the paramagnetic region, $\rho(T)$ forms a broad hump at about 100~K
and exhibits a negative logarithmic slope below about 20~K, both manifesting the Kondo scattering
of conduction electrons on crystalline electric field $4f$ sublevels (cf. \cite{kaczorowski}). A
drop of the resistivity at about 6~K results from the antiferromagnetic phase transition. For the
alloys with $x>0$, the magnetic phase transition could not be evidenced because of the temperature
limit of our resistivity measurements ($\approx 4$~K). With increasing the nickel content, the low
temperature $-\log{T}$ contribution becomes dominant and for $x=0.4$ it entirely merges with the
hump at elevated temperatures. Over an extended temperature range $\rho(T)$ of this alloy exhibits
a single nearly logarithmic slope [\fref{fig_resist}(a)]. For higher $x$, the low temperature
resistivity evolves into a $T^2$ dependence [see dashed curve in \fref{fig_resist}(b)], that
signals the intermediate valence behaviour.

In order to shed more light on the observed evolution of $\rho (T)$ of the alloys
Ce$_2$Cu$_{1-x}$Ni$_{x}$In with $x \leqslant 0.4$, we derived the magnetic contribution $\rho_{\rm
mag}$ to their total resistivity by subtracting the phonon contribution to $\rho(T)$ of
La$_2$Cu$_2$In (cf. \cite{kaczorowski}). The $\rho_{\rm mag} (T)$ curve obtained for Ce$_2$Cu$_2$In
revealed that the hump in $\rho (T)$ results from the presence of a distinct maximum in $\rho_{\rm
mag}$ followed by another $-\log{T}$ dependence at high temperatures [\fref{fig_resist_magn}(a)].
Moreover, two logarithmic slopes are visible also in the other Cu-rich alloys, in which the maximum
at 100~K is somewhat obscured by the low temperature $-\log{T}$ dependence.

According to the Cornut--Coqblin model~\cite{cornut}, in the localized regime with the magnetic
ground state the interplay of the Kondo effect and the crystalline electric field (CEF) results in
the magnetic contribution to the total resistivity, which can be expressed as:
\begin{equation}
\label{cornut}
\rho_{\rm i}(T) = a J_{\rm ex}^2 \frac{\lambda^2_i -1}{\lambda_i (2J_1+1)}
+ 2 a N(E_{\rm F}) J_{\rm ex}^3 \frac{\lambda^2_i -1}{(2J_1+1)} \log{\frac{T}{D_i}},
\end{equation}
where $J_1 \! = \! 5/2$, $\lambda_i$ denotes the effective degeneracy of the $4f^1$ state, $J_{\rm
ex}$ is the negative exchange integral, $N(E_{\rm F})$ is the density of states at the Fermi level,
$D_i$ is the effective cutoff parameter, and $a$ is a constant. In the case of Ce$^{3+}$ ions
experiencing a tetragonal crystal electric field potential, the sixfold degenerated $4f^1$
multiplet splits into three doublets located at the energies $\Delta_1 \equiv 0 < \Delta_2 <
\Delta_3$. Population of these levels depends on the temperature, so the effective degeneracy of
the $4f^1$ state increases with increasing $T$ from $\lambda_1$~=~2 (ground doublet) at absolute
zero temperature, to $\lambda_2$~=~4 (pseudoquartet) at elevated $T$, to $\lambda_3$~=~6
(pseudosextet) at high temperatures. From \eref{cornut} it becomes clear that the slope and the
magnitude of $\rho_i(T)$ depend on $\lambda_i$, so for well separated doublets one can observe up
to three $-\log{T}$ slopes in the magnetic contribution to the resistivity.

However, from the point of view of the present study, it is more important to note that the
relative magnitudes of the particular $\rho_i(T)$ contributions [equation \eref{cornut}] may vary
also with $J_{\rm ex}$, making the observation of the separated Kondo slopes difficult (cf.
\cite{cornut}). Assuming Boltzman-like thermal excitations of the crystalline electric field
doublets (in line with the Cornut--Coqblin approach~\cite{cornut}), the Kondo contribution
$\rho_{\rm mag}$ to the electrical resistivity can be expressed as:
\begin{equation}
\label{cornut_simulation}
\rho_{\rm mag}(T) = \rho_1 + (\rho_2-\rho_1)\exp{\left(-\frac{\Delta_2}{T}\right)}
+ (\rho_3-\rho_2)\exp{\left(-\frac{\Delta_3}{T}\right)}.
\end{equation}
Note, that the contributions $\rho_2$ and $\rho_3$ include already $\rho_1$ and $\rho_2$,
respectively. In order to avoid their multiplication, two corrections, i.e.
$-\rho_1\exp{(-\Delta_2/T)}$ and $-\rho_2\exp{(-\Delta_3/T})$, were added. In order to check the
influence of magnetic exchange on the overall character of $\rho_{\rm mag}(T)$, we made a few
simulations using a fixed crystal electric field scheme (2:2:2 with $\Delta_1$~=~0~K,
$\Delta_2$~=~50~K, and $\Delta_2$~=~100~K) and varying the arbitrary chosen value of $J_{\rm ex}$.
For simplicity, the constants $a$, $N(E_{\rm F})$, and $D_{\rm i}$ were assumed to be independent
of $J_{\rm ex}$ and equal to 1. \Fref{cornut_simulation}(b), \fref{cornut_simulation}(c) and
\fref{cornut_simulation}(d) display the results of these calculations.

As shown in \fref{cornut_simulation}(b), for low absolute values of $J_{\rm ex}$ the magnitude of
the ground doublet contribution $\rho_1$ is much less than the pseudo-quartet ($\rho_2$) and
pseudo-sextet ($\rho_3$) contributions. As a consequence, the crossover between the crystal field
levels manifests itself as a distinct maximum in $\rho_{\rm mag}(T)$. Upon increasing $|J_{\rm
ex}|$, $\rho_1$ increases and becomes of the same order of magnitude as $\rho_2$ and $\rho_3$, and
the maximum in $\rho_{\rm mag}(T)$ becomes less pronounced [\fref{fig_resist_magn}(c)]. Finally,
for large $|J_{\rm ex}|$, the magnitude of $\rho_1$ is higher than $\rho_2$ and $\rho_3$, so the
crossover between the logarithmic slopes in $\rho_{\rm mag}(T)$ is monotonic
[\fref{fig_resist_magn}(d)].

Although any quantitative analysis of the resistivity data in terms of \eref{cornut_simulation} is
hampered by polycrystalline nature of the specimens studied (effects of magnetocrystalline
anisotropy, inter-grains resistance, possible internal cracks, etc.), it is worth emphasizing the
qualitative similarity of the experimental [\fref{fig_resist_magn}(a)] and model curves
[\fref{fig_resist_magn}(b), \fref{fig_resist_magn}(c) and \fref{fig_resist_magn}(d)]. Their
evolution with increasing the Ni content can thus be attributed to the increase in the strength of
the hybridization, as it has been done for the magnetic susceptibility data.

\section{Summary and conclusions}

The partial substitution of Cu by Ni in the antiferromagnetically ordered compound Ce$_2$Cu$_2$In
results in a monotonic compression of its unit cell (up to 10\%), accompanied by a drastic change
of the physical properties of the system. In particular, the magnetic measurements of
Ce$_2$Cu$_{2-x}$Ni$_x$In revealed continuous evolution from the localised magnetic moments regime
for $x \lesssim 0.6$ to the fluctuating valence state for $x \gtrsim 1.6$. The detailed analysis of
the inverse magnetic susceptibility data within the ICF model yielded reversal of the order of the
magnetic $4f^1$ and nonmagnetic $4f^0$ levels in the alloys with $0.6 \lesssim x \lesssim 1.6$. In
other words, the $4f^1$ level, being the ground state in the Cu-rich alloys and lying {\it below}
the Fermi level (equivalent to the $4f^0$ configuration), becomes an excited state in the Ni-rich
alloys, lying {\it above} the Fermi level.

Deeply in the localized regime, the Ce magnetic moments order antiferromagnetically, and the
N\'{e}el temperature decreases quasi-linearly with increasing the nickel content from 5.5~K in
Ce$_2$Cu$_2$In down to 2.7~K in Ce$_2$Cu$_{1.2}$Ni$_{0.8}$In. Simultaneous significant broadening
of the maximum in $\chi(T)$ at $T_{\rm N}$ indicates possible smearing of the phase transition at
temperatures approaching the absolute zero. The analysis of the temperature dependencies of the
electrical resistivity of the Ce$_2$Cu$_{2-x}$Ni$_x$In series revealed an evolution of the system
from the dense Kondo lattice regime towards the nonmagnetic metal regime, controlled mainly by the
increase in the exchange integral.

All the presented findings are in line with the Doniach picture of Kondo systems, which predicts a
change in the character of the ground state from long range magnetic order to intermediate valence.
This transformation is induced by an increase in the value of the exchange integral $J_{\rm ex}$
that measures the hybridization strength. For Ce$_2$Cu$_{2-x}$Ni$_x$In, the increase of $J_{\rm
ex}$ with increasing $x$ is likely a consequence of the observed compression of the unit cell
volume. Simultaneous reduction in a number of the $3d$ electrons, which could probably lead to some
decrease in $J_{\rm ex}$, seems to have less significant influence on the physical properties of
the system. To verify our hypothesis detailed spectroscopic studies of the Ce$_2$Cu$_{2-x}$Ni$_x$In
series are indispensable.

\ack
This work was supported by the Polish Ministry of Science and Higher Education within research
grant no. N N202 102338.

\end{document}